\documentclass[12pt,a4paper]{article}

\usepackage{epsfig}
\usepackage{exscale}
\def\lsim{\raise0.3ex\hbox{$<$\kern-0.75em\raise-1.1ex\hbox{$\sim$}}}
\def\gsim{\raise0.3ex\hbox{$>$\kern-0.75em\raise-1.1ex\hbox{$\sim$}}}
\newcommand{\<}{\langle}
\renewcommand{\>}{\rangle}
\newcommand{\be}{\begin{equation}}
\newcommand{\ee}{\end{equation}}
\newcommand{\ba}{\begin{eqnarray}}
\newcommand{\ea}{\end{eqnarray}}

\def\spose#1{\hbox to 0pt{#1\hss}}
\def\ltapprox{\mathrel{\spose{\lower 3pt\hbox{$\mathchar"218$}}
 \raise 2.0pt\hbox{$\mathchar"13C$}}}
\def\gtapprox{\mathrel{\spose{\lower 3pt\hbox{$\mathchar"218$}}
 \raise 2.0pt\hbox{$\mathchar"13E$}}}
\def\phv{\vec \phi}

\def\ad#1{$\,^{\rm #1}$}
\def\NT{N_\tau}
\def\nt{\ifmmode\NT\else$\NT$\fi}
\def\NS{N_\sigma}
\def\ns{\ifmmode\NS\else$\NS$\fi}

\def\p{^\prime}
\def\v{\vec}
\def\n{\noindent}

\begin{document}
\begin{titlepage} 
\thispagestyle{empty}

 \mbox{} \hfill BI-TP 2011/06\\
 \mbox{} \hfill hep-lat/1105.0584v2\\
 \mbox{} \hfill February 2012
\begin{center}
\vspace*{0.8cm}
{{\Large \bf The scaling functions of the free energy density\\
and its derivatives for the \boldmath$3d~O(4)$ model\\}}\vspace*{1.0cm}
{\large J.\ Engels\ad 1 and F.\ Karsch\ad {1,2}}\\ \vspace*{0.8cm}
\centerline {{$^{\rm 1}$}{\em Fakult\"at f\"ur Physik, 
    Universit\"at Bielefeld, D-33615 Bielefeld, Germany}} \vspace*{0.4cm}
\centerline {{\large $^{\rm 2}$}{\em Physics Department Brookhaven 
National Laboratory, Upton, NY 11973}}
\protect\date \\ \vspace*{0.9cm}
{\bf   Abstract   \\ } \end{center} \indent
We derive direct representations of the scaling functions
of the $3d~O(4)$ model which are relevant for comparisons to other
models, in particular QCD. This is done in terms of expansions in the
scaling variable $z=\bar t/h^{1/\Delta}$. The expansions around $z=0$ 
and the corresponding asymptotic ones for $z \rightarrow \pm\infty\,$ 
overlap such that no interpolation is needed.\,
The expansion coefficients are determined numerically from the data 
of a previous high statistics simulation of the $O(4)$ model
on a three-dimensional lattice of linear extension $L=120$. From the
scaling function of the magnetization we calculate the leading asymptotic
coefficients of the scaling function of the free energy density. As
a result we obtain the universal amplitude ratio $A^+/A^-=1.84(4)$ for
the specific heat. Comparing the scaling function of the energy density 
to the data we find the non-singular part of the energy density 
$\epsilon_{ns}(T)$ with high precision and at the same time excellent 
scaling properties.

\vfill \begin{flushleft} 
PACS : 64.10.+h; 75.10.Hk; 05.50+q \\ 
Keywords: Scaling function; $O(4)$ model; Goldstone modes;
free energy density\\ 
\noindent{\rule[-.3cm]{5cm}{.02cm}} \\
\vspace*{0.2cm}
E-mail addresses: engels@physik.uni-bielefeld.de, 
karsch@physik.uni-bielefeld.de, karsch@bnl.gov 
\end{flushleft} 
\end{titlepage}


\section{Introduction}

The aim of the paper is to provide representations of the scaling functions
of the three-dimensional $O(4)$ model which can be used in tests of other
models on their membership of the corresponding universality class. This
is especially of importance for quantum chromodynamics (QCD) with two 
degenerate light-quark flavours at finite temperature, because it is
believed \cite{Pisarski:ms}-\cite{Ejiri:2009ac} to belong to the $O(4)$
universality class at its chiral transition in the continuum limit.
There exist already many parametrizations
\cite{Toussaint:1996qr}-\cite{Braun:2007td} of the magnetic equation of 
state. They differ essentially in the following aspects:
\begin{itemize} 
\item[a)] The form of the magnetic equation of state which is 
initially parametrized. The most used form is the Widom-Griffiths (WG) 
form \cite{Widom:1965,Griffiths:1967}, where both the scaling function
and the scaling variable depend on the magnetization $M\,$. 
A second form has the advantage that the scaling variable 
is independent of $M\,$. It is therefore more appropriate for 
the comparison to $M$-data with errors. The two forms are 
completely equivalent in describing the critical behaviour
of the model and they can be derived from each other. In
principle, it is then only necessary to parametrize one form. 
\item[b)] The type of parametrization which is used. The
parametrization has to describe the correct general scaling
laws as deduced from renormalization group (RG) theory, it must 
satisfy Griffiths's analyticity conditions \cite{Griffiths:1967}
and take into account the Goldstone singularities in the low
temperature phase.
\item[c)] The input from which the parameters are determined. 
There are two main sources of information: from field-theory methods, 
for example $\epsilon$-expansions, small-field expansions or high 
temperature series etc.\ , and secondly, from Monte Carlo (MC) data.
\end{itemize}
Our paper is inspired by the pioneering work of Toussaint
\cite{Toussaint:1996qr}. In his paper the scaling function (of the 
second kind) for the order parameter was calculated based only on MC 
data and moreover these data were simulated at finite 
external field $H\,$. The main purpose of the paper was, like
ours, to provide the scaling function for the QCD analysis. 
The parametrization was carried out in a third, unusual 
form, which has not been used since then, yet the Goldstone
effects had still not been taken into account. Like the 
WG form, this third form has the drawback, that the 
calculation of the second form of the scaling function and
in particular its derivatives with respect to its scaling
variable is an indirect one and it is therefore unhandy. 
Moreover, scaling functions of the second kind determine via
the location of their extrema the important pseudocritical lines. 
In Refs.\ \cite{Engels:1999wf} and \cite{Engels:2003nq} the WG form
was used and parametrized with a combination of a low and a high 
temperature ansatz in accord with the requirements of b). The two
parts were subsequently connected by an interpolation. The 
parameters were deduced exclusively from MC data with finite external 
fields. In Refs.\ \cite{ParisenToldin:2003hq,Cucchieri:2004xg} and 
\cite{Braun:2007td} the WG form was used. All three papers discuss
and use variants of the classical parametric representation of the 
equation of state introduced by Schofield and Josephson 
\cite{Schofield:1969zza}-\cite{Schofield:1969zz} in 1969, which is 
valid in the whole critical region. Refs.\ \cite{ParisenToldin:2003hq}
and \cite{Cucchieri:2004xg} differ in details of the parametrization
and in the input: whereas \cite{ParisenToldin:2003hq} relies 
essentially on field theory input and uses the data for testing, 
Ref.\ \cite{Cucchieri:2004xg} determines its parameters directly
from fits to the data. In Ref.\ \cite{Braun:2007td} the functional
RG method is used to calculate the scaling functions. The classical
representation had been invented for the WG form, at a time where
only few MC data on small lattices were available. Today it is still 
used for all kind of calculations of universal quantities from 
field theory. However, as we shall demonstrate, it is not necessary  
to work with this representation. Instead we parametrize directly 
the scaling functions of the second kind, which is the preferred 
choice of the QCD community in their tests on the universality class 
of the $O(4)$ model. In addition, this allows us not only to make use
of the magnetization data in the determination of the parameters,
but also of the data for the susceptibilities.

In order to broaden the tests on the universality class we calculate as
well the scaling functions connected to the energy density and the 
specific heat. Here however, the initially unknown regular or
non-singular part of the energy density and/or specific heat is contained
in the data and has to be subtracted correctly. We show how this can be
done in principle, when the critical exponent $\alpha$ is negative. The
inadequate estimate of the regular parts in a former test for 2-flavour
QCD \cite{D'Elia:2005bv} prevented a successful outcome of the test from 
the beginning and led to unjustified conclusions. Yet, there is a 
derivative of the energy density, the thermal susceptibility or 
covariance between the energy density and the magnetization, which does
not require a subtraction of the non-singular term. We shall use also
this quantity for our parametrization. 
  
The specific model which we study here is the standard $O(4)$-invariant
nonlinear $\sigma$-model, which is defined by
\be
\beta\,{\cal H}\;=\;-J \,\sum_{<{\vec x},{\vec y}>}\phv_{\vec x}\cdot
\phv_{\vec y} \;-\; {\vec H}\cdot\,\sum_{{\vec x}} \phv_{\vec x} \;,
\ee
where ${\vec x}$ and ${\vec y}$ are nearest-neighbour sites on a 
three-dimensional hypercubic lattice, and $\phv_{\vec x}$ is a
four-component unit vector at site ${\vec x}$. The coupling $J$ and
the external magnetic field $\v H$ are reduced quantities, that is
they contain already a factor $\beta=1/T$. In fact, we consider in
the following the coupling directly as the inverse temperature,
$J\equiv 1/T$. 

It is useful to decompose 
the spin vector $\phv_{\vec x}$ into longitudinal (parallel to the magnetic 
field $\vec H$) and transverse components 
\be
\phv_{\vec x}\; =\; \phi_{\vec x}^{\parallel} \vec e_H +
\phv_{\vec x}^{\perp} ~,\quad {\rm with}\quad \vec e_H=\vec H/ H~,
\ee
where $H$ is the magnitude of the magnetic field. We define the energy
of a spin configuration as
\be
E\;=\;- \sum_{<{\vec x},{\vec y}>}\phv_{\vec x}\cdot \phv_{\vec y}\;.
\ee
The lattice average $\phi^{\parallel}$ of the longitudinal spin components
is
\be
\phi^{\parallel}  \;=\; \frac{1}{V}\sum_{{\vec x}} \phi_{\vec x}^{\parallel}
~,
\ee 
where $V=L^3$ and $L$ is the number of lattice points per direction.
The partition function is then
\be
Z(T,H)\;=\; \int \prod_{\v x} d^{\,4}\phi_{\v x}\;\delta (\phv_{\v x}^{\,2}
 -1) \exp(-\beta\,E + HV\phi^{\parallel} ) ~.
\ee
We introduce the (reduced) free energy density as usual by 
\be
f(T,H) \;=\; -\frac{1}{V}\ln Z ~, 
\ee
from which one obtains the order parameter of the system, the 
magnetization $M$, as
\be
M \;=\; - \frac{\partial f}{\partial H}\; =\;\<\, \phi^{\parallel} \,\>~.
\label{Mterm}
\ee
The longitudinal susceptibility is the second derivative of $-f$
with respect to the field
\be
\chi_L \;=\; {\partial M \over \partial H}
 \;=\; V(\<\, \phi^{\parallel2} \,\>-M^2)~. 
\label{chil}
\ee
The energy density is
\be
\epsilon \;=\;  \frac{\partial f}{\partial \beta}\; =\;
\frac{\<\, E \,\>}{V}~,
\label{epsi}
\ee
and the specific heat
\be
C \;=\;  \frac{\partial \epsilon}{\partial T}\; =\;
\frac{\beta^2}{V}(\<\, E^2 \,\>-\<\, E \,\>^2)~.
\label{spech}
\ee
Finally we define the thermal susceptibility $\chi_t$ as the mixed second 
derivative of $f$
\be
\chi_t \;=\; {\partial M \over \partial \beta}
 \;=\; \<\, E \,\>\< \, \phi^{\parallel} \,\>-\<\,E \phi^{\parallel} \,\>~.
 \label{chit}
\ee
The rest of the paper is organized as follows. First we discuss the
critical behaviour of the observables and the universal scaling 
functions, which we want to calculate. In Section 3 we describe 
the expansions with which we parametrize the scaling functions. 
Some details of the used simulations and the parametrizations resulting
from the data are presented in Section 4. Here we also investigate the
r\^ole of the non-singular terms for the scaling of the data.
We close with a summary and the conclusions.


\section{Critical behaviour and scaling functions}
\label{section:Criti}

In the thermodynamic limit ($V\rightarrow \infty$) the above defined
observables show power law behaviour close to $T_c$. It is described by
critical amplitudes and exponents of the reduced temperature 
$t=(T-T_c)/T_c$ for $H=0$ and the magnetic field $H\,$ for $t=0\,$,
respectively. According to RG theory the non-analytic or singular part 
$f_s$ of the free energy density is responsible for critical behaviour. 
Besides $f_s\,$, the free energy density contains a regular or
non-singular part $f_{ns}$. Correspondingly, the derivatives of $f_{ns}$
contribute regular terms to the scaling laws, which apart from the cases
of the energy density and the specific heat (for $\alpha<0$) are sub-leading. 
In the two-dimensional Ising model such an analytic contribution to the 
magnetic susceptibility was established \cite{Caselle:2001jv,Orrick:2001}.
In Ref.\ \cite{Toussaint:1996qr}, Toussaint makes a corresponding
ansatz $f_{ns}=c_{H2}H^2+c_{J1}t+c_{J2}t^2+c_{J3}t^3$, which leads to an
additional constant in $\chi_L$, a term $\sim H$ in $M$ and an 
$H$-independent $\epsilon_{ns}(T)$. Since in our former scaling fits to
$M$ at $T_c$, e.\ g.\ in Refs.\ \cite{Engels:2003nq} and 
\cite{Engels:2009tv}, we never discovered such a regular term we follow 
Privman et al.\ \cite{Privman:1991} and assume the non-singular part 
$f_{ns}$ to have no field dependence, that is 
\be
 f(T,H)\;=\; f_s(T,H) + f_{ns}(T)~,\quad {\rm and}
\ee 
\be
\epsilon\;=\; \epsilon_s + \epsilon_{ns}(T)~,\quad C\;=\; C_s+C_{ns}(T)~.  
\ee 
The regular parts do not disappear at $T=T_c\,$. We may expand
$\epsilon_{ns}(T)$ in $T$ at $T_c$ 
\be
\epsilon_{ns}(T) \;=\; \epsilon_{ns}(T_c) + (T-T_c)\cdot C_{ns}(T_c) +
\frac{1}{2}(T-T_c)^2 \cdot C^{\;\prime}_{ns}(T_c) +\dots~.
\label{Tayeps}
\ee
\n The scaling laws at $H=0$ are then for
the magnetization (from now on $\beta$ denotes a critical exponent)
\be
 M  \;=\; B (-t)^{\beta} \quad {\rm for~} t<0~,
\label{mcr}
\ee
the longitudinal susceptibility
\be
 \chi_L \;=\; C^{+} t^{-\gamma} \quad {\rm for~} t > 0~,
\label{chicr}
\ee 
and the energy density and the specfic heat both for $t<0$ and $t>0$
\be
\epsilon \;=\;  \epsilon_{ns}(T) + \frac{A^{\pm}}{\alpha(1-\alpha)}
 T_c t|t|^{-\alpha}~,
\label{epscr}
\ee
\be
C \;=\; C_{ns}(T) +  \frac{A^{\pm}}{\alpha}|t|^{-\alpha}~.
\label{ccr}
\ee
For the thermal susceptibility we have for $t<0$
\be
\chi_t \;=\; \beta B T_c  (-t)^{\beta-1} ~. 
\label{chitcr}
\ee
\n On the critical line $T=T_c$ or $t=0$ we have for $H>0$ 
the scaling laws
\be
M \;=\; B^cH^{1/\delta} \quad {\rm or}\quad H \;=\;D_c M^{\delta}~,
\label{mcrh}
\ee
\be
\chi_L \;=\; C^cH^{1/\delta-1} \quad {\rm with}\quad C^c \;=\;B^c/\delta~.
\label{chicrh}
\ee
The remaining observables scale as follows
\be
\epsilon \;=\; \epsilon_{ns}(T_c) + E_c H^{(1-\alpha)/\Delta}~,
\label{epscrh}
\ee
\be
C_s-\frac{2\epsilon_s}{T_c} \;=\;
\frac{A_c}{\alpha_c} H^{-\alpha_c}~,
\label{ccrh}
\ee
\be
\chi_t \;=\; X_c H^{(\beta-1)/\Delta}~,
\label{chitcrh}
\ee
where $\alpha_c=\alpha/\Delta$ and $\Delta=\beta\delta$
is the so-called "gap exponent".

\n Generalizations of these scaling laws to both non-zero $t$ and 
$H$-values may be derived from the RG scaling equation for $f_s$
\be
f_s(u_1,u_2,u_3,\dots) \;=\; b^{-d} f_s(b^{y_1} u_1,b^{y_2} u_2,
b^{y_3} u_3,\dots)~. 
\label{RGse}
\ee
Here, the $u_j$ with $j=1,2,\dots$ are the scaling fields, $b$ is a
positive scale factor and the $y_j$ are the RG eigenvalues. 
The class of our model has two relevant scaling fields 
$u_1=u_t,~u_2=u_h$ with $y_t,y_h>0$ and infinitely many irrelevant ones 
with negative $y_j$. The relevant scaling fields depend analytically
on $t$ and $H$ and 
\be
u_t\;=\; c_tt+O(t^2,H^2)~,\quad u_h \;=\; c_hH +O(tH)~.
\ee
The $c_t,c_h$ are two model-dependent (positive) metric scale factors.
Choosing $b=u_h^{-1/y_h}$ for $H>0$ one obtains from Eq.\ (\ref{RGse})
the second form of scaling functions from
\be
f_s(u_t,u_h,u_{j>2})=u_h^{d/y_h}f_s(u_tu_h^{-y_t/y_h},1,u_ju_h^{-y_j/y_h})~.
\ee 
Close to the critical point, for $t,H$ small, $u_t=c_tt,~u_h=c_hH\,$,
and the dependence on the irrelevant scaling fields becomes negligible,
$f_s$ is a universal scaling function of $u_t$ and $u_h\,$ and
\be
f_s\;=\; (c_hH)^{d/y_h}\Psi_2(c_tc_h^{-y_t/y_h}tH^{-y_t/y_h})~,
\label{fspsi}
\ee
where $\Psi_2$ is again a universal function. By comparison with the 
scaling laws one obtains
\be
y_t\;=\;1/\nu\;,\; y_h\;=\;1/\nu_c\;=\;\Delta/\nu\;,\;{\rm or~}\;
\Delta\;=\;y_h/y_t~,
\ee
and the hyperscaling relations 
\be
2-\alpha \;=\; d\nu, \quad \gamma \;=\; \beta (\delta -1), \quad
d\nu \;=\; \beta (1 +\delta)~.
\label{hyps}
\ee
Instead of working with two metric scale factors one usually introduces 
new temperature and field variables $\bar t= tT_c/T_0$ and $h=H/H_0$
which are chosen such that the scaling laws for the magnetization
simplify to 
\be
M(t=0) = h^{1/\delta} \quad {\rm and } \quad H_0 = D_c~,
\label{normh}
\ee
\be
M(h=0) = (-\bar t\;)^{\beta} \quad {\rm and } \quad T_0 = B^{-1/\beta}T_c~.
\label{normt}
\ee
The magnetic equation of state as derived from Eqs.\ (\ref{Mterm}) 
and (\ref{fspsi}) becomes then
\be
M\;=\;h^{1/\delta} f_G(z)~,
\label{mscale}
\ee
where $f_G$ is a universal scaling function with the argument
\be
 z \;=\; \bar t/h^{1/\Delta}~.
\label{zdef}
\ee
It fulfills the normalization conditions
\be
f_G(0)\; =\; 1~,\quad {\rm and}\quad f_G(z) {\raisebox{-1ex}{$
\stackrel{\displaystyle\longrightarrow}{\scriptstyle z 
\rightarrow -\infty}$}}(-z)^{\beta}~.
\label{normfg}
\ee
Due to Eq.\ (\ref{Mterm}), the corresponding scaling equation 
of the free energy density must then be
\be
f_s \;=\; H_0 h^{1+1/\delta} f_f(z)~,
\label{fsscale}
\ee
where $f_f(z)$ is again a universal scaling function, and
\be
f_G(z) \;=\; -\left(1+\frac{1}{\delta}\right) f_f(z)
+\frac{z}{\Delta}f_f\p(z)~.
\label{fgdiff}
\ee

\n Since the susceptibility $\chi_L$ is the derivative of $M$ with
respect to $H$ we obtain from Eq.\ (\ref{mscale})
\be
\chi_L = {\partial M \over \partial H} = {h^{1/\delta -1} \over H_0}
 f_{\chi}(z)~,
\label{cscale}
\ee
with 
\be
f_{\chi}(z) = {1 \over \delta} \left( f_G(z) - {z \over \beta} f_G\p (z)
\right)~.
\label{fchi}
\ee
For $H\to 0$ at fixed $t>0$, that is for $z\rightarrow \infty$, the 
leading asymptotic term of $f_{\chi}$ is determined by Eq.\ (\ref{chicr})
\be
f_{\chi} (z)\; {\raisebox{-1ex}{$\stackrel 
{\displaystyle =}{\scriptstyle z \rightarrow \infty}$}}
\;  C^+ D_c B^{\delta-1} z^{-\gamma}\;=\;R_{\chi} z^{-\gamma} ~,
\label{fcasy}
\ee
where $R_{\chi}$ is a universal amplitude product. For $z\rightarrow \infty$
the leading terms of $f_G$ and $f_{\chi}$ are identical, because
for $T>T_c$ and small magnetic field $M \propto H\,$. The rest of our
observables are related to $f_f(z)$ and $f_G(z)$ as follows
\ba
\epsilon_s  \!\!& =&\!\! -\frac{T^2}{T_0}H_0 h^{(1-\alpha)/\Delta}
f_f\p(z)~, \label{epsiscale}\\
C_s-\frac{2}{T}\epsilon_s \!\!& =&\!\!-\left(\frac{T}{T_0}\right)^2
H_0 h^{-\alpha/\Delta} f_f''(z)~, \label{spechscale}\\
\chi_t \!\!& =&\!\! -\frac{T^2}{T_0} h^{(\beta-1)/\Delta} f_G\p(z)~.
\label{chitscale}
\ea

\section{Expansions of the scaling functions}
\label{section:expand}

\n In principle we have to parametrize only one scaling function, either
$f_f(z)$ or $f_G(z)\,$, because they are related by the differential
equation (\ref{fgdiff}). We choose as usual $f_G(z)$, because it is 
directly calculable from the magnetization data. Our representation of the
scaling function is composed of three expansions: one around $z=0$ and
two for $z\rightarrow \pm \infty\,$. In the following we derive relations
between the expansion coefficients of $f_f(z)$ and $f_G(z)\,$. We start
with the expansions for small $z$
\be
f_f(z)\;=\; \sum_{n=0}^\infty a_nz^n~,\quad f_G(z)\;=\; 
\sum_{n=0}^\infty b_nz^n~.
\label{sum0}
\ee
From Eq.\ (\ref{fgdiff}) we obtain
\be
b_n\;=\; \left[ -\left(1+\frac{1}{\delta}\right)+\frac{n}{\Delta} \right]
 a_n~,~{\rm or}\quad a_n\;=\;\frac{\Delta b_n}{\alpha+n-2}~.
\ee
The last equation connects the derivatives of the two scaling functions
at $z=0$ 
\be
f_f^{(n)}(0)\;=\; \frac{\Delta}{\alpha+n-2}\cdot f_G^{(n)}(0)~,
\label{f(n)(0)}
\ee
and because of the first of the normalization conditions, Eq.\ 
(\ref{normfg})
\be
f_G(0)=b_0\equiv 1~,\quad{\rm and}\quad f_f(0)=a_0=\frac{\Delta}{\alpha-2}~.
\label{f(0)}
\ee
Next we consider the asymptotic expansion in the high temperature region,
that is for $z\rightarrow \infty$, or for $t>0$ and $h\rightarrow 0$. 
Since $M$ is an odd function of $H$ for $t>0$ (Griffiths's condition),
we must have 
\be
f_G(z)\;=\; z^{-\gamma} \cdot \sum_{n=0}^\infty d_n^+ z^{-2n\Delta}~,
\label{fGas+}
\ee
The prefactor is the leading term of $f_G$ and $d_0^+=R_\chi$ (see Eq.\ 
(\ref{fcasy}) and the remark after it). The corresponding ansatz for 
$f_f(z)$ is
\be
f_f(z)\;=\; z^{2-\alpha} \cdot \sum_{n=0}^\infty c_n^+ z^{-2n\Delta}~,
\label{ffas+}
\ee
Using again Eq.\ (\ref{fgdiff}) we are led to the relation
\be
c_{n+1}^+\;=\; \frac{-d_n^+}{2(n+1)}~,\quad {\rm with}\quad 
c_1^+\;=\;-\,\frac{R_{\chi}}{2}~,
\label{cn+}
\ee
however, the coefficient $c_0^+$ is not specified by the last equation.

\n In the low temperature region, for $t<0$ and $h\rightarrow 0$, that is
for $z\rightarrow -\infty$, massless Goldstone modes appear. They lead to
a divergence of the transverse susceptibility $\chi_T\sim H^{-1}$. In
addition also the longitudinal susceptibility $\chi_L$ is diverging on the
coexistence curve. Here, the predicted divergence in three dimensions is 
\cite{Wallace:1975}-\cite{Anishetty:1995kj}
\be
\chi_L (T<T_c,H) \sim H^{-1/2}~.
\label{chigold}
\ee
This is equivalent to a dependence of the magnetization on $H^{1/2}$ near
the coexistence curve \cite{Engels:1999wf}. Therefore we make the 
following ansatz for $f_G(z)$ in this region
\be
f_G(z)\;=\; (-z)^{\beta} \cdot \sum_{n=0}^\infty d_n^- (-z)^{-n\Delta/2}~,
\label{fGas-}
\ee
where due to the second normalization condition in Eq.\ (\ref{normfg}),~ 
$d_0^-=1\,$. The corresponding ansatz for $f_f(z)$ is
\be
f_f(z)\;=\; (-z)^{2-\alpha} \cdot \sum_{n=0}^\infty c_n^-(-z)^{-n\Delta/2}~.
\label{ffas-}
\ee 
Inserting the two expansions into the differential equation (\ref{fgdiff})
we find
\be
c_{n+2}^- \;=\; -\frac{2d_n^-}{n+2}~.
\label{cn-}
\ee
As in the high temperature phase the coefficient of the leading term,
$c_0^-$, is not fixed, moreover $c_1^-\equiv 0$, and $c_2^-=-1\,$.
In order to completely solve Eq.\ (\ref{fgdiff}) for $f_f(z)$ we still 
have to find the coefficients $c_0^{\pm}$. Since $\alpha<0$ for the
$O(4)$ model we may proceed in the following way. First we consider
the small $z$-expansions for $z>0$ 
\ba
 \sum_{n=3}^\infty a_nz^n  \!\!& =&\!\! \Delta \sum_{n=3}^\infty 
\frac{b_n z^n}{\alpha+n-2} ~\!=\!~ \Delta z^{2-\alpha}
\int_0^z dy y^{\alpha-3}\sum_{n=3}^\infty b_n y^n\\
 \!\!& =&\!\! \Delta z^{2-\alpha}\int_0^z dy y^{\alpha-3}
\left[ f_G(y)-1-b_1y-b_2y^2 \right]\\
 \!\!& =&\!\! f_f(z) -a_0-a_1z-a_2z^2~.
\ea  
That enables us to calculate $c_0^+$
\be
c_0^+\;=\; \lim_{z\rightarrow\infty} f_f(z)z^{\alpha-2}\;=\;
 \Delta \int_0^\infty dy y^{\alpha-3}\left[ f_G(y)-1-b_1y-b_2y^2 \right]~,
\label{c_0+a}
\ee
or, by partial integration
\be
c_0^+\;=\;\frac{\Delta}{2-\alpha} \int_0^\infty dy y^{\alpha-2}
\left[ f_G\p(y)-f_G\p(0)-yf_G''(0) \right]~.
\label{c_0+b}
\ee
In the same manner we can calculate $c_0^-$ from an integral over negative
$z$ by starting from
\be
f_f(z)\;=\; a_0+a_1z+a_2z^2+\Delta(-z)^{2-\alpha}\int_z^0 dy(-y)^{\alpha-3}
\left[ f_G(y)-1-b_1y-b_2y^2 \right]\,,
\ee
and taking the limit $z\rightarrow -\infty$
\be
c_0^-\;=\; \lim_{z\rightarrow- \infty} f_f(z)(-z)^{\alpha-2}\;=\;
 \Delta \int^0_{-\infty} dy(-y)^{\alpha-3}\left[ f_G(y)-1-b_1y-b_2y^2 \right]~,
\label{c_0-a}
\ee
or
\be
c_0^-\;=\;\frac{-\Delta}{2-\alpha} \int^0_{-\infty} dy(-y)^{\alpha-2}
\left[ f_G\p(y)-f_G\p(0)-yf_G''(0) \right]~.
\label{c_0-b}
\ee
The function $f_f(z)$ is universal as a whole and so are each of its
expansion coefficients, that is $c_0^+$ and $c_0^-$ are universal. In
fact, it can be shown, that these coefficients are the universal 
products of critical amplitudes
\be
c_0^\pm \;=\; f_s^\pm (B^c)^\delta B^{-(1+\delta)}~,
\label{c0uni}
\ee 
where
\be
f_s^\pm \;=\; \frac{A^\pm}{-\alpha(1-\alpha)(2-\alpha)}~,
\label{fspm} 
\ee
are the critical amplitudes of the free energy density for $H=0$
and $t\ne 0$
\be
f \;=\; f_{ns}(T) + f_s^\pm |t|^{2-\alpha}~.
\label{fcr}
\ee
Because of Eqs.\ (\ref{c0uni}) and (\ref{fspm}) we can now calculate the
universal ratio 
\be
\frac{A^+}{A^-} \;=\; \frac{c_0^+}{c_0^-} 
\label{U0}
\ee
from the two integrals in Eqs.\ (\ref{c_0+b}) and (\ref{c_0-b}). A similar
formula is known for the magnetic equation of state in the Widom-Griffiths
form \cite{Aharony:1974}. It was for example used to determine the ratio
$A^+/A^-$ in the case of the $O(2)$ model \cite{Cucchieri:2002hu}.  

\section{The parametrizations}
\label{section:results}

\n The data we use in the following to parametrize the scaling functions
were all obtained from simulations described in detail in Ref.\
\cite{Engels:2009tv}. We repeat here only the main features of these
simulations. They were performed on three-dimensional lattices with 
periodic boundary conditions and linear extension $L=120$. The coupling
constant region which was explored is $0.90 \le J \le 1.2$, the magnetic
field was varied from $H=0.0001$ to $H=0.007\,$. In general 100000 
measurements were done at each fixed $H$ and $J$. We have reevaluated the
raw data to obtain the magnetization, the longitudinal and the thermal
susceptibilities, the energy density and the specific heat. 
Due to the large spatial volume of the lattice, most of the 
finite size effects have disappeared from the data. This is in 
particular true for the energy density and the magnetization,
to a smaller extent also for the susceptibilities and the
specific heat. We shall discuss the remaining effects when the
scaling of the respective observables is analyzed. A further
source of difficulties is, at larger $|t|$ and $H$-values, the 
possible appearance of corrections to scaling because of the
influence of irrelevant scaling fields. These violations of 
scaling should be visible in the scaling plots for the data.
However, as we shall show below, we find perfect scaling 
properties without any sign of these corrections to scaling 
for our values of $t$ and $H$ for the energy density and the 
magnetization. This not the case for the corresponding non-linear 
$O(2)$ model, see e.\ g.\ Fig.\ 5 of Ref.\ \cite{Engels:2000xw},
where strong scaling violations were found for $T<T_c$. An 
explanation for this striking difference can be inferred from two 
papers, by Hasenbusch and T\"or\"ok \cite{Hasenbusch:1999} for $N=2$
and by Hasenbusch \cite{Hasenbusch:2000ph} for $N=4$. In these papers 
the leading corrections to scaling could be eliminated by using 
instead of the non-linear $O(N)$ invariant models the 
corresponding $O(N)$ symmetric $\phi^4$ models and the tuning of
the additional parameter $\lambda\,$. It turned out that the 
optimal parameter value is 2.1 for $N=2$ and 12.5(4.0) for $N=4$. 
The non-linear case corresponds to $\lambda=\infty$. One expects 
therefore to find significantly weaker corrections for $N=4$ as 
compared to $N=2$. The remarkable lack of scaling corrections 
had already been noted in Ref.\ \cite{Engels:1999wf}, where for
the first time the critical exponents for the $3d$ $O(4)$ model
were determined from magnetization data at finite external fields.
Later, in Ref.\ \cite{Engels:2003nq}, the result for the exponent
$\delta$ could be improved with better data and fits where
correction-to-scaling terms had been taken into account. Yet,
these terms were contributing at best marginally and could as
well be neglected.

In order to define our variables $t,\bar t,h$ and $z$ we use the same
critical amplitudes, temperature and exponent values as in Ref.\ 
\cite{Engels:2003nq} and \cite{Engels:2009tv}. These are
\be
J_c=T_c^{-1}=0.93590~,\quad T_0=1.093~,\quad H_0=4.845~,
\label{def1}
\ee
\be
\beta=0.380~,\quad \delta=4.824~,\quad \Delta=1.83312~,
\label{def2}
\ee
and from the hyperscaling relations
\be
\alpha=-0.2131~,\quad \nu=0.7377~,\quad \gamma=1.4531~.
\label{def3}
\ee
We have compared the exponents from Eqs.\ (\ref{def2}) and 
(\ref{def3}) to the field theory results of Guida and 
Zinn-Justin \cite{Guida:1998}, displayed in their Table 3. Apart
from the value for $\delta=4.824(9)$ (from \cite{Engels:2003nq})
which corresponds to the value $\eta =0.0302(16)$ 
via the hyperscaling relation $\eta= (5-\delta)/(1+\delta)\,$
our numbers are always close to the respective central values of
Guida and Zinn-Justin and they are always inside their error bars.
 
We procede in the following way. First we calulate $f_G(z)$ from the
magnetization data and fit the large $z$ parts to the asymptotic
expansions. The small $z$ region is more intricate, because the 
derivative $-f_G\p(z)$ has a peak for $z>0$, which determines the
pseudocritical line. In order to model the corresponding variation
properly we fit directly the derivative for small $z$. It can be
obtained either from the data for $\chi_t$, Eq.\ (\ref{chitscale}),
or from $\chi_L$ and $M$, using Eqs.\ (\ref{mscale}), (\ref{cscale}) and 
(\ref{fchi}). After completion of the parametrization of $f_G(z)$
we compute the leading asymptotic coefficients $c_0^{\pm}$ of $f_f(z)$.
The scaling function of the free energy density is then also entirely
known. In the next step we determine the non-singular contributions
to the energy density and the specific heat, first at $T_c$ and then
at all our $T$-values. We show that the results for $\epsilon_{ns}$
shape a smooth function of $T$, where $T_c$ is not a distinguished
point. With this function $\epsilon_{ns}(T)$ it is then possible to
compare the scaling functions to the data for the energy density and
specific heat.

\begin{figure}[t]
\begin{center}
   \epsfig{bbllx=105,bblly=70,bburx=690,bbury=488,
       file=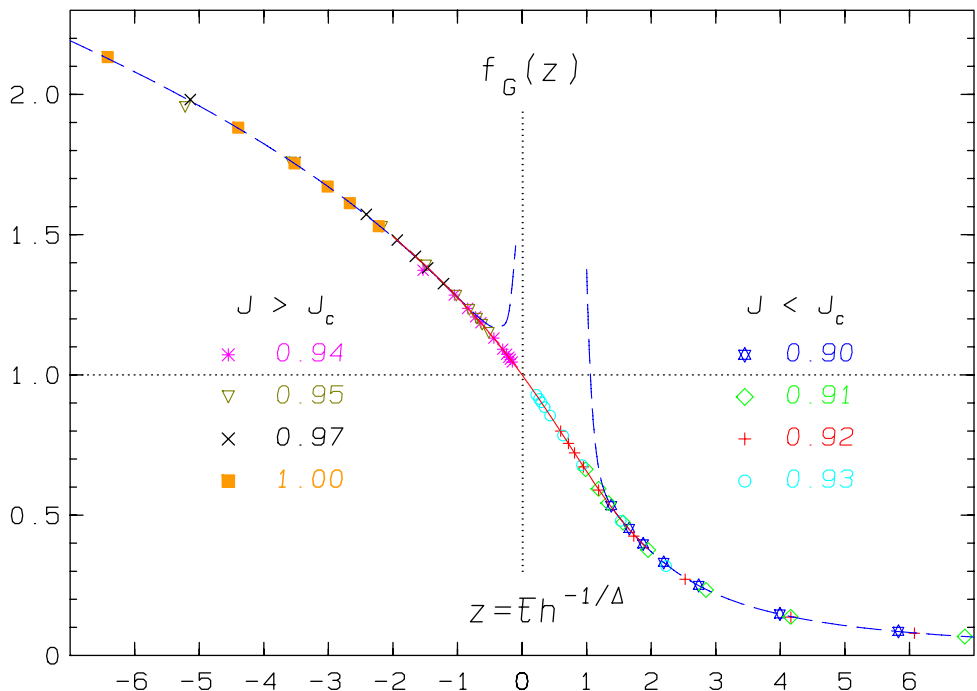, width=120mm}
\end{center}
\caption{The scaling function $f_G=Mh^{-1/\delta}$ as a function of 
$z=\bar t h^{-1/\Delta}$. The dashed lines show the asymptotic 
expansions, the solid line the Taylor expansion at $z=0$. The
numbers refer to the different $J=1/T$-values of the data.}
\label{fig:f_G}
\end{figure}
In Fig.\ \ref{fig:f_G} we show the data obtained from the magnetization
for the scaling function $f_G(z)$ and our parametrizations. Obviously
the data scale very well, apart 
from the data for $H=0.0001$ and $z<0$, which show some finite size
effect (in the figure at $z=-1.534\,$ and $-5.219\,$). We have fitted
$f_G$ in the asymptotic regions with the first three terms of the 
respective expansions from Eqs.\ (\ref{fGas+}) and (\ref{fGas-}). For
the positive $z$-range $[1.5,15]$ we found the coefficients
\be
d_0^+=1.10599\pm 0.00555~,~~d_1^+=-1.31829\pm 0.1087~,~~
d_2^+=1.5884\pm 0.4646~.
\label{fitas+}
\ee
In the negative $z$-range $[-10,-1]$ we discarded the data with
$H=0.0001$ and obtained
\be
d_0^-\equiv 1~,~~d_1^-=0.273651\pm 0.002933~,~~
d_2^-=0.0036058\pm 0.004875~.
\label{fitas-}
\ee
\clearpage
\begin{figure}[t]
\begin{center}
   \epsfig{bbllx=103,bblly=68,bburx=689,bbury=486,
       file=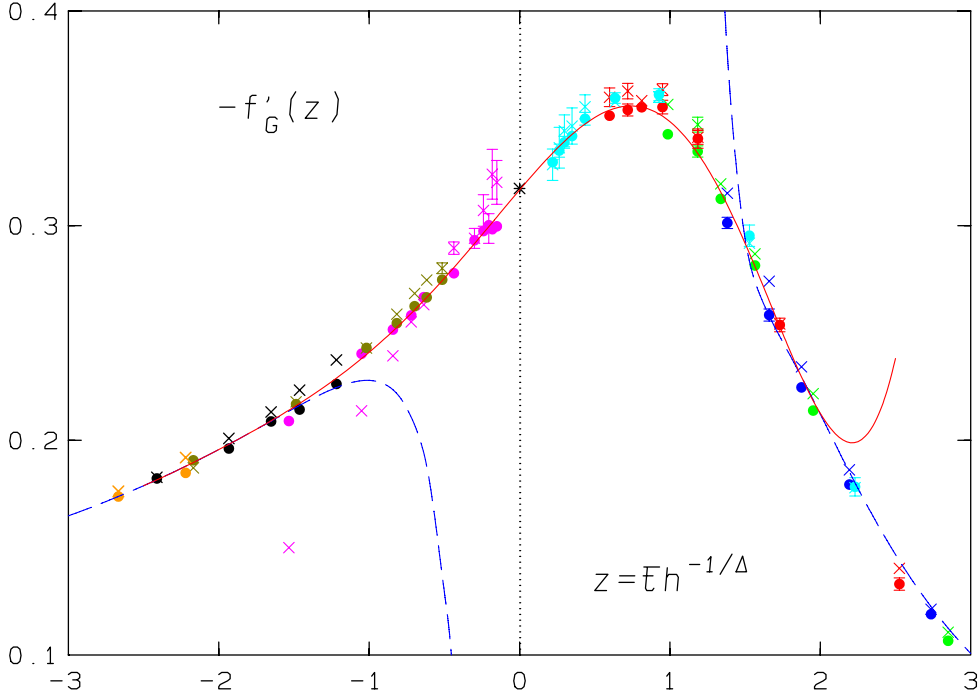, width=120mm}
\end{center}
\caption{The derivative $-f_G\p(z)$ as a function of 
$z=\bar t h^{-1/\Delta}$. The filled circles denote the data calculated
from $\chi_t$, the crosses the data obtained from $\chi_L$ and $M\,$.
The dashed lines show the asymptotic expansions, the solid line the Taylor
expansion around $z=0$.}
\label{fig:f_Gpr}
\end{figure}
\n Since $d_0^+=R_{\chi}$ we have a new value for this quantity, which
is compatible with the old values $R_{\chi}=1.084(18)$ from Ref.\ 
\cite{Engels:2003nq} and $R_{\chi}=1.12(11)$ from Ref.\ 
\cite{ParisenToldin:2003hq} but somewhat more accurate. Astonishingly,
the asymptotic expansions describe the function $f_G(z)$ very well
down to rather small $|z|$-values, and as can be seen in Fig.\ 
\ref{fig:f_G} they overlap with our approximation to the Taylor 
expansion at $z=0\,$. As mentioned already, we use the derivative
$-f_G\p(z)$ to determine the coefficients of the Taylor expansion.
In Fig.\ \ref{fig:f_Gpr} we show the data which we obtained from
$\chi_t$ (filled circles) and $\chi_L$ and $M$ (crosses) for the 
derivative. Obviously, the data involving $\chi_L$
suffer from large finite size effects in the whole low temperature
region ($z<0$) for already moderately small $H$-values. This behaviour
is known and a consequence of Eq.\ (\ref{chigold}), the divergence of
$\chi_L$ near the coexistence line. In contrast to that, the data from 
$\chi_t$ show a consistent scaling behaviour for $z<0$ (apart from the
$H=0.0001$ point for $J=0.94$ at $z=-1.534\,$). For $z>0$ but close to
the critical point we find still larger finite size effects for the 
$\chi_L$-data as compared to the ones for the $\chi_t$-data, because of
the stronger divergence of $\chi_L$ on the critical line. At larger 
positive $z$-values beyond the peak region we observe however, that the 
$\chi_t$-data are systematically smaller than the $\chi_L$-data. The 
reason for this lies in the cluster update \cite{Wolff:1988uh} which
was used to produce the data. That update diminishes very efficiently
the autocorrelation time for the order parameter, but is less efficient
for the energy density. For increasing temperature and/or decreasing $H$
the cluster size drops and if the number of cluster updates is not
correspondingly increased the 
autocorrelation times for the energy density data increase faster 
than those for the magnetization data. An increasing decorrelation of 
$\epsilon$ and $M$ takes then place with increasing $z$, that is the 
$\chi_t$-data become too small. In view of all these considerations we
use in our Taylor fits only $\chi_t$-data for $z<1.3$ and in the 
$z$-interval $[1.3,2]$ both types of data. 

\begin{figure}[t]
\begin{center}
   \epsfig{bbllx=102,bblly=109,bburx=689,bbury=443,
       file=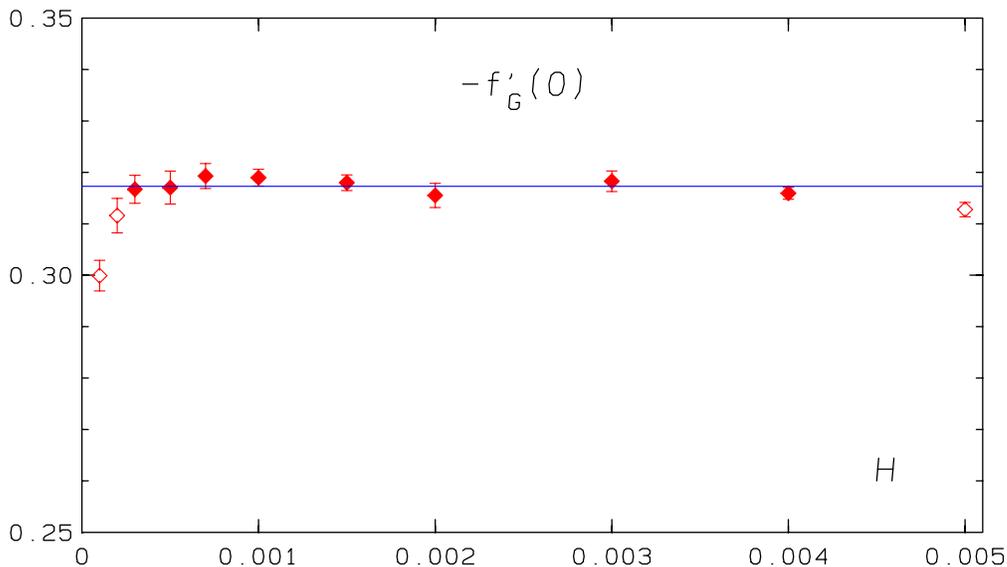, width=120mm}
\end{center}
\caption{The quantity $-f_G\p(0)$ as a function of the magnetic field 
$H$. The filled data points were used for the fit, the solid line shows
the fit result.}
\label{fig:fg01}
\end{figure}

From our data at $T_c$ we have calculated an additional data point at 
$z=0$, denoted by a star in Fig.\ \ref{fig:f_Gpr}. To this end we have
used Eq.\ (\ref{chitscale}) at $T_c$
\be
-f_G\p(0) = \frac{T_0}{T_c^2}h^{-(\beta-1)/\Delta}\chi_t(T_c)~.
\label{chitjc}
\ee 
The corresponding data are shown in Fig.\ \ref{fig:fg01}. At small $H$
we have again finite size effects, at large $H$ corrections to scaling
set in. A fit to the remaining data points leads to the result
$-f_G\p(0)=0.3173(5)$.  
We have approximated the Taylor series with polynomials. Instead of 
using a single very high order polynomial, we started with two fits to
6. order polynomials, that is for $f_G\p(z)$ with the ansatz 
\be
f_G\p(z) = b_1 +2b_2z+3b_3z^2+4b_4z^3+5b_5z^4+6b_6z^5~,
\label{fitfGpr} 
\ee
in the overlapping $z$-intervals $[-2.5,0.75]$ and $[-0.75,2]$. The 
lowest coefficients of the two polynomials should, at the end of 
the process, coincide to generate a smooth parametrization close to
$z=0$. In the second step we took therefore the arithmetic averages
of the results of the two fits for $b_1$ and $b_2$, fixed them and 
repeated the two fits to determine the remaining coefficients. In step
three $b_3$ was fixed by averaging the corresponding means of the first
and the second step. The last fits were performed with fixed $b_1,b_2$
and $b_3$ in the intervals $[-2.5,0.75]$ and $[-0.5,2]$, including some
points from the fits to the asymptotic regions. Our final result is
\ba
& b_0\equiv 1~,~~b_1=-0.3166125\pm 0.000534~,
\label{b01} \\
& b_2=-0.04112553\pm 0.001290~,~~b_3=~0.00384019\pm 0.000667~. 
\label{b23}
\ea 
The remaining coefficients are different for $z<0$ and $z>0$. We find
for $z>0$
\ba 
 & b_4^+=~0.006705475\pm 0.001704~,~~b_5^+=~0.0047342\pm 0.001429~,
\label{b45}\\
 & b_6^+=-0.001931267\pm 0.000312~, 
\label{b6}
\ea
and for $z<0$
\ba 
 & b_4^-=~0.007100450\pm 0.000160~,~~b_5^-=~0.0023729\pm 0.000095~,
\label{b45-}\\
 & b_6^-=~0.000272312\pm 0.000021~. 
\label{b6-}
\ea
We note that $b_4^-$ and $b_4^+$ still coincide inside their
error bars. In Fig.\ \ref{fig:f_Gpr} we have plotted the respective
approximations to $-f_G\p(z)$ in the $z$-ranges $[-2.5,0]$ and 
$[0,2.5]$. Obviously, there is a large range for $z<0$ and a shorter 
range for $z>0$ where the approximations overlap and coincide with
the respective asymptotic expansions. In the rest of the paper we use
therefore the Taylor expansions in the $z$-range $[-2,1.95]$ and 
outside the asymptotic expressions.

It is now straightforward to calculate the coefficients of the leading
asymptotic terms of $f_f(z)$ from Eqs.\ (\ref{c_0+b}) and (\ref{c_0-b}).
We find
\be
 c_0^+= 0.422059886\pm 0.010595~,\quad c_0^- = 0.229176194\pm 0.010669~.
\label{c0pm}
\ee
The errors of $c_0^\pm$ have been determined using the complete
correlation matrix of the contributing parameters. The main contributions
to $c_0^\pm$ are coming from the two terms proportional to $b_1$ and
$b_2$. The second, larger term is the same for both coefficients, the
first only changes the sign, that is for $c_0^+$ we have the sum, for 
$c_0^-$ the difference of these terms and as a consequence the $c_0^\pm$
are strongly correlated. That allows us to estimate the correlation
between $c_0^+$ and $c_0^-$ to $C_{+-}=-\sigma_1^2+\sigma_2^2$, where
the $\sigma_{1,2}$ are the errors of the two terms. From Eqs.\ (\ref{U0})
and (\ref{c0pm}) we obtain then the universal ratio
\be
\frac{A^+}{A^-}= 1.842 \pm 0.043~.
\label{A+-}
\ee 
Our value for the ratio is in agreement with the final estimates found
in Refs.\ \cite{ParisenToldin:2003hq}, 1.91(10), and
\cite{Cucchieri:2004xg}, 1.8(2)\,. We note however, that our error 
estimate does not include a possible variation of the critical 
exponents used. 
\begin{figure}[b]
\begin{center}
   \epsfig{bbllx=102,bblly=81,bburx=688,bbury=472,
       file=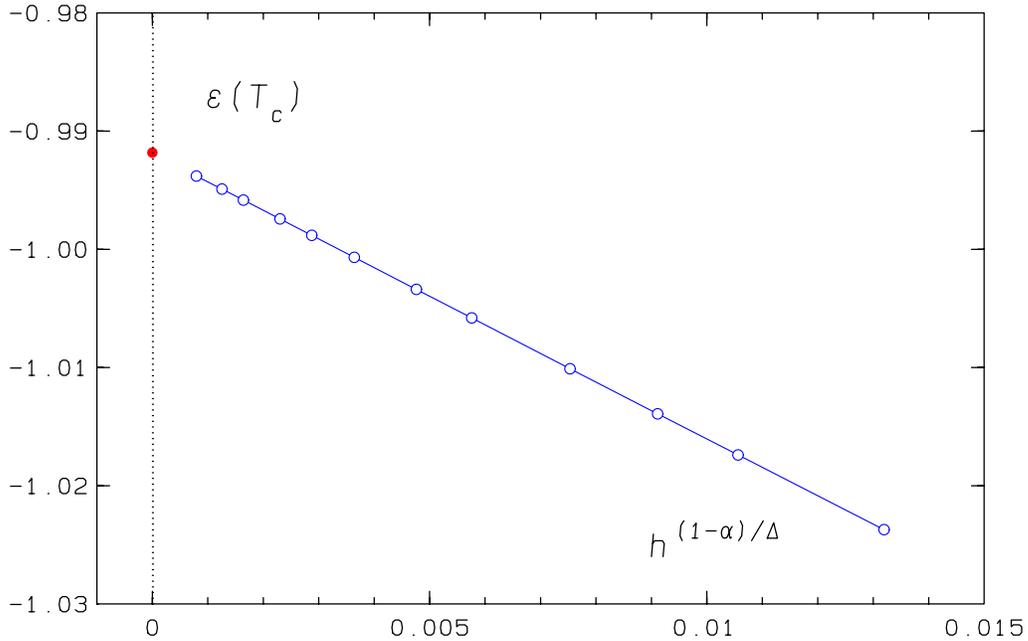, width=118mm}
\end{center}
\caption{The energy density at $T_c$ (open circles) as a function of  
$h^{(1-\alpha)/\Delta}\,$. The filled circle shows
the extrapolation to $h=0$, that is $\epsilon_{ns}(T_c)$, the points
are connected by straight lines.}
\label{fig:epsitc}
\end{figure}

With the results from Eq.\ (\ref{c0pm}) we have completely specified 
the parametrizations of $f_G(z)$ and of $f_f(z)$. In order to test the
scaling function of the free energy density and/or its temperature
derivatives we still need the regular contributions. Quite generally,
the exponent $\alpha$ must be always less than 1 for continuous 
transitions, because the energy density at $T_c$ is finite and there is
no latent heat (which would be possible for $\alpha=1$). From Eqs.\
(\ref{epscr}) and (\ref{epscrh}) we know therefore that  
\be
\epsilon_s(T_c,H=0)=0~,\quad {\rm or}\quad 
\epsilon(T_c,H=0)=\epsilon_{ns}(T_c)~.
\label{epsattc}
\ee
We may then determine $\epsilon_{ns}(T_c)$ from our data for the energy 
density at $T_c$. In Fig.\ \ref{fig:epsitc}\break we see that the data 
at $T=T_c$ fulfill the expected scaling law, Eq.\ (\ref{epscrh}) and
Eq.\ (\ref{epsiscale}) accordingly
\be
\epsilon_s(T_c)= -\frac{T_c^2}{T_0}H_0 h^{(1-\alpha)/\Delta}
f_f\p(0)~.
\label{epsis}
\ee
Moreover, there is no sign of a finite size dependence at small external
fields.
We have fitted the data directly with Eq.\ (\ref{epscrh}) and find
\be
\epsilon_{ns}(T_c)=-0.991888(13)~,~~E_c=-0.8500(06)~,~~
f_f\p(0)=0.47723(32)~.
\label{tceps2}
\ee 
Evidently we have a very precise result for the non-singular part of the
energy density
\n at $T_c$. The derivative $f_f\p(0)$ can also be calculated from 
$b_1=f_G\p(0)$ and Eq.\ (\ref{f(n)(0)}). That leads to 
$f_f\p(0)=0.47843(81)$ and is consistent with the previous result. 

\begin{figure}[t]
\begin{center}
   \epsfig{bbllx=96,bblly=81,bburx=683,bbury=472,
       file=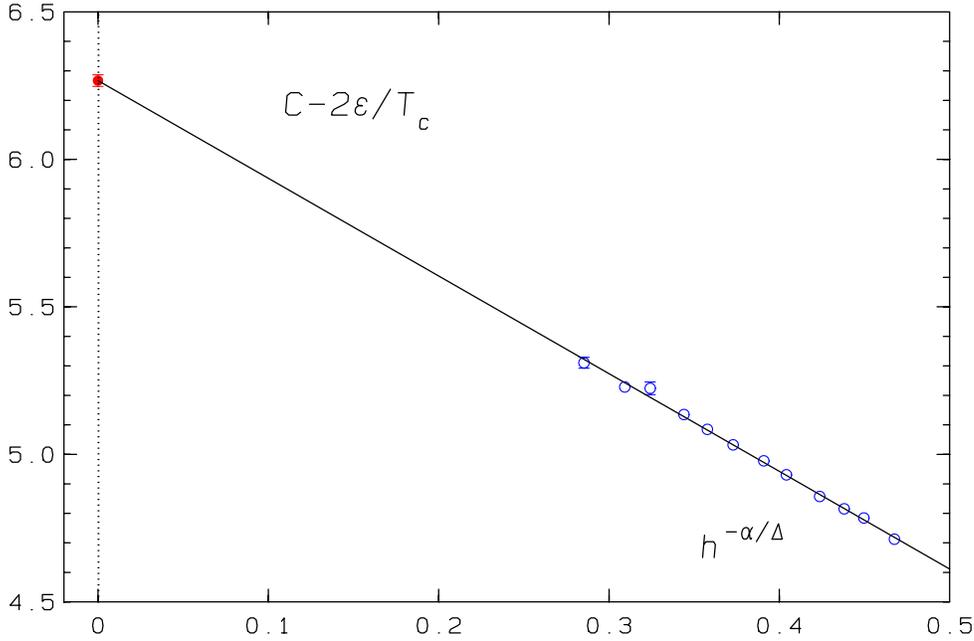, width=118mm}
\end{center}
\caption{The quantity $C-2\epsilon/T$ at $T_c$ (open circles) as a
function of $h^{-\alpha/\Delta}\,$. The filled circle shows the 
extrapolation to $h=0$, the straight line the fit,
Eq.\ (\ref{spechscale}).}
\label{fig:cme}
\end{figure}

For positive $\alpha$ the specific heat is diverging and it is therefore
unclear how to determine the regular term $C_{ns}(T_c)$. If however
$\alpha$ is negative, then $C_s-2\epsilon_s/T$ disappears at $T_c$ for 
$h=0$, because of Eq.\ (\ref{ccrh}), the finite regular term remains
and we can calculate $C_{ns}(T_c)$. Usually, the specific heat has
nevertheless a sharp peak at $T_c$ and $H=0$, which just means that the
critical amplitudes are negative. That sign is taken care of by the
factors $1/\alpha$ and $1/\alpha_c\,$, respectively in our amplitude
definitions, so that $A^\pm$ and $A_c$ are positive.
In Fig.\ \ref{fig:cme} we show our data for $C-2\epsilon/T$ at $T_c$
and the corresponding fit to Eqs.\ (\ref{ccrh}) and (\ref{spechscale}).
We find 
\be
\left(C_{ns}-\frac{2\epsilon_{ns}}{T}\right)(T_c) = 6.2669(195)~,~
{\rm and}~~C_{ns}(T_c) = 4.4103(195)~,
\label{cmeattc}
\ee
\be
A_c = 0.32041(468)~,~{\rm or}~~ f_f''(0)= 0.7151(104)~,
\label{Ac}
\ee
which is compatible to the value $f_f''(0)= 0.7075(221)$ calculated
from $b_2\,$. 

\begin{figure}[t!]
\begin{center}
   \epsfig{bbllx=99,bblly=68,bburx=685,bbury=486,
       file=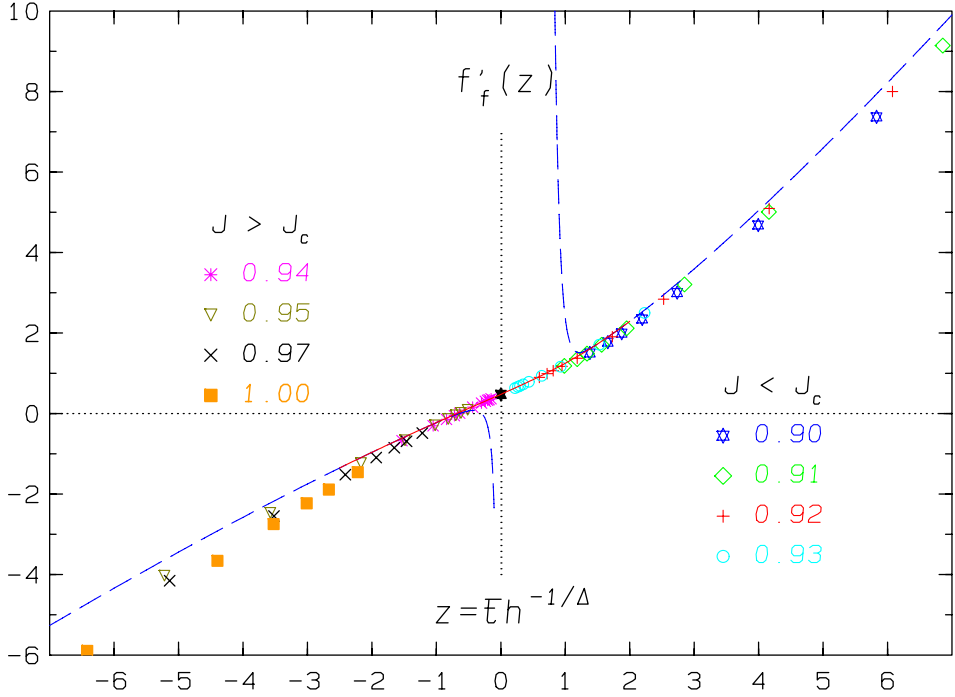, width=120mm}
\end{center}
\caption{The scaling function $f_f\p(z)$ as a function of 
$z=\bar t h^{-1/\Delta}$. The dashed lines show the asymptotic expansions, 
the solid line the polynomial approximations for small $z$. The star at 
$z=0$ is the result from Eq.\ (\ref{tceps2}).}
\label{fig:ff1a}
\end{figure}

\begin{figure}[t!]
\begin{center}
   \epsfig{bbllx=111,bblly=96,bburx=696,bbury=462,
       file=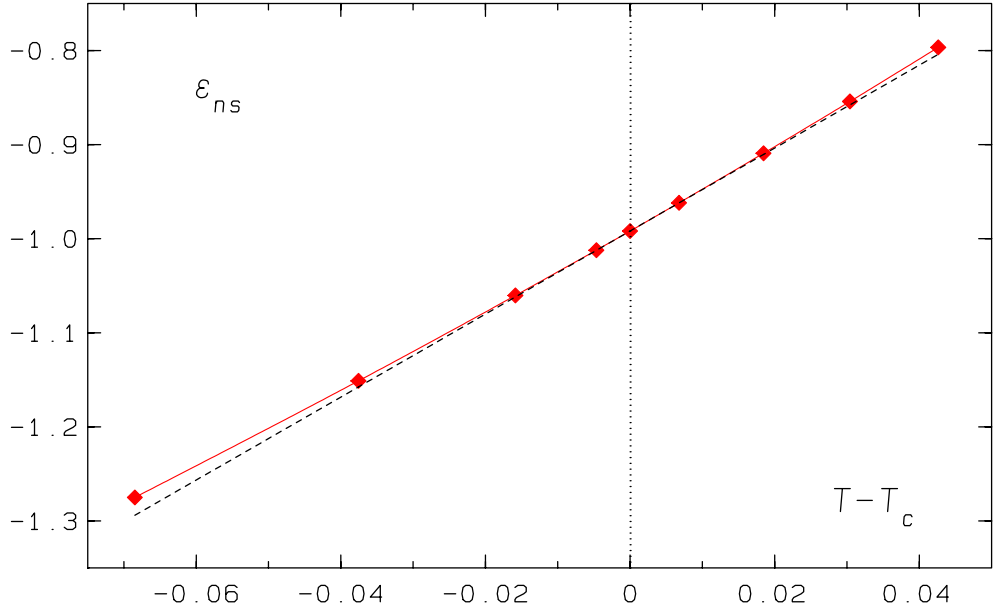, width=120mm}
\end{center}
\caption{The non-singular part of the energy density $\epsilon_{ns}(T)$
as a function of $T-T_c$ (filled diamonds). The dashed line 
shows the approximation from Eq.\ (\ref{ensa}).}
\label{fig:epsns}
\end{figure}

\begin{figure}[b]
\begin{center}
   \epsfig{bbllx=99,bblly=68,bburx=685,bbury=486,
       file=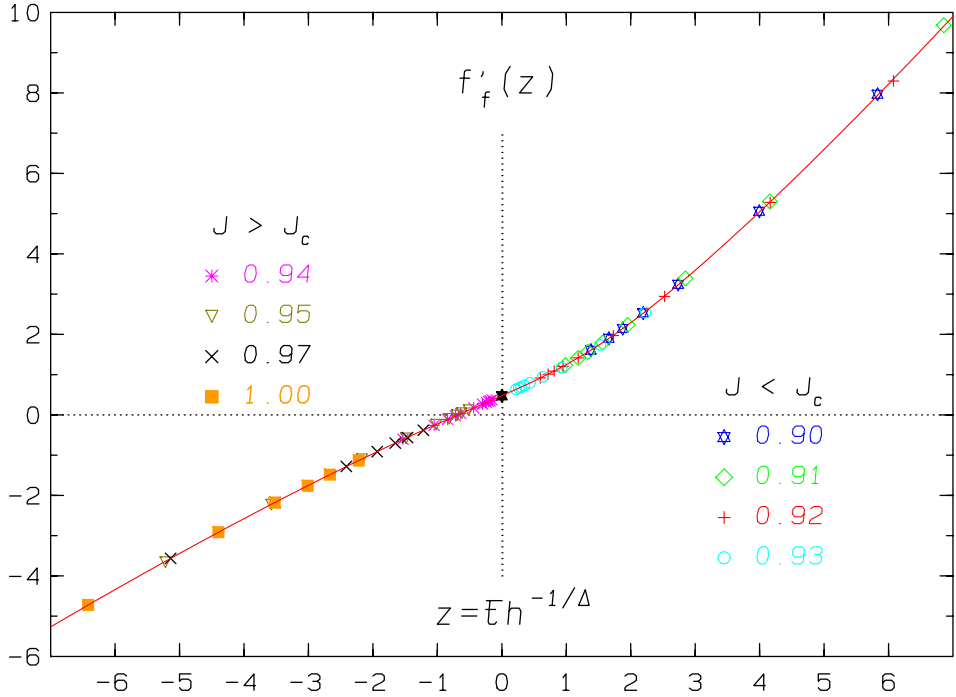, width=120mm}
\end{center}
\caption{The scaling function $f_f\p(z)$ as a function of 
$z=\bar t h^{-1/\Delta}$. The line shows our parametrization, the data
have been calculated using Eqs.\ (\ref{Tay01}) and (\ref{Tay23}).}
\label{fig:ff1_2}
\end{figure}

In Fig.\ \ref{fig:ff1a} we show the scaling function $f_f\p(z)$ which
is obtained from our para\-metri\-zation. We plot the
asymptotic expansions and the Taylor expansion separately and find
that they are overlapping in the same regions as for $f_G(z)$, that is 
our calculation of the $c_0^{\pm}$ is consistent. The data which we also
show have been calculated assuming that
\be
\epsilon_{ns}(T) = \epsilon_{ns}(T_c) + (T-T_c)\cdot C_{ns}(T_c)~,
\label{ensa}  
\ee
\n where the numbers are from our fits at $T=T_c$. We note here, that
without the term proportional to $C_{ns}(T_c)$ the data would scale
nowhere apart from the point $z=0$. In Fig.\ \ref{fig:ff1a} we observe
scaling for small $|z|$ and coincidence with the predicted scaling
function, but not at larger values. The reason for that is the
assumption, Eq.\ (\ref{ensa}), for the function $\epsilon_{ns}(T)$,
which leads to inaccurate $\epsilon_{ns}$-values for larger $|T-T_c|$.
We can test this and the scaling behaviour at fixed $T$ and varying $H$
at the same time. If we have scaling then Eq.\ (\ref{epsiscale}) must
hold and 
\be
\epsilon_{ns}(T)= \epsilon(T,H) +T^2 \frac{H_0}{T_0}
h^{(1-\alpha)/\Delta}f_f\p(z)~,
\label{ensT}
\ee
where the $\epsilon(T,H)$ are the energy density data at fixed $T$ and 
$f_f\p(z)$ is the predicted scaling function. The test is successful, 
if we obtain the same value $\epsilon_{ns}(T)$ for all $H$ inside the
error bars. This is indeed the case. The errors of the averages are
tiny. They vary for $T\le T_c$ between $6\cdot 10^{-6}$ and 
$1.4\cdot 10^{-5}$, and in the hot phase they increase to 
$3.5\cdot 10^{-5}$. In Fig.\ \ref{fig:epsns} we compare the found
values for $\epsilon_{ns}(T)$ with the approximation from
Eq.\ (\ref{ensa}), where we have used our previously calculated 
numbers from Eqs.\ (\ref{tceps2}) and (\ref{cmeattc}). Though the 
differences in Fig.\ \ref{fig:epsns} do not seem to be large, they are
the reason for the deviation of the data from the scaling function 
$f_f\p(z)$ in Fig.\ \ref{fig:ff1a}\,. We have fitted our results for
$\epsilon_{ns}(T)$ with the Taylor expansion, Eq.\ (\ref{Tayeps}), up 
to the third derivative of $\epsilon_{ns}$ and find
\ba
\epsilon_{ns}(T_c) \!\!\!\!& =&\!\!\!\! -0.991792(28)~,
\quad C_{ns}(T_c) = 4.3910(14)~, \label{Tay01}\\ 
C_{ns}'(T_c)  \!\!\!\!& =&\!\! 8.448(108)~,~~
\quad C_{ns}''(T_c) = 42.79\pm 5.13~. \label{Tay23}
\ea
The new result for $C_{ns}(T_c)$ is in agreement with the one we
obtained from our admittedly far extrapolation of $C-2\epsilon/T$
in $h$ at $T_c\,$. We have used the results from the Taylor expansion fit
to approximate the functions $\epsilon_{ns}(T)$ and $C_{ns}(T)$ in the
calculation of the scaling functions $f_f'$ and $f_f''$ from our data. 
As can be seen from Fig.\ \ref{fig:ff1_2} we find now perfect scaling 
for the energy density, even for relatively
\clearpage
\begin{figure}[ht]
\begin{center}
   \epsfig{bbllx=105,bblly=70,bburx=690,bbury=488,
       file=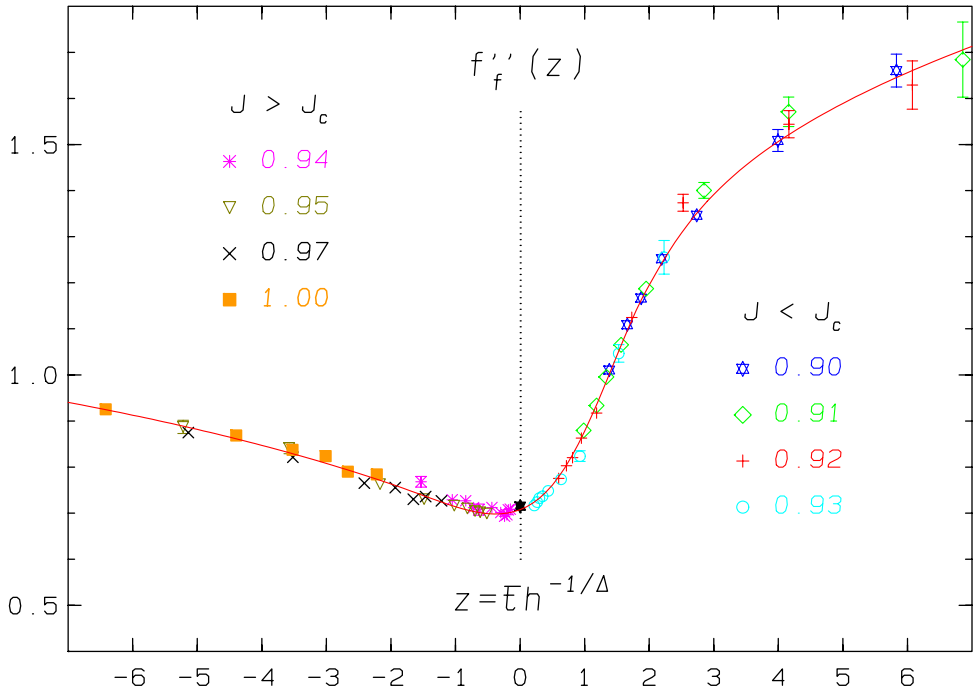, width=120mm}
\end{center}
\caption{The scaling function $f_f''(z)$ as a function of 
$z=\bar t h^{-1/\Delta}$. The line shows our parametrization, the data
have been calculated using Eqs.\ (\ref{Tay01}) and (\ref{Tay23}),
the star at $z=0$ is the result from Eq.\ (\ref{Ac}).}
\label{fig:ff2_2}
\end{figure}
\n large $|T-T_c|$-values.
In Fig.\ \ref{fig:ff2_2} we compare our parametrization for $f_f''$
with the data. Since the specific heat is proportional to the fluctuation
of the energy density, the data are not as precise as for $f_f'$, 
especially for high temperatures. Nevertheless we observe satisfactory 
scaling and a further confirmation for our parametrization.

\begin{figure}[t!]
\begin{center}
   \epsfig{bbllx=46,bblly=23,bburx=585,bbury=409,
       file=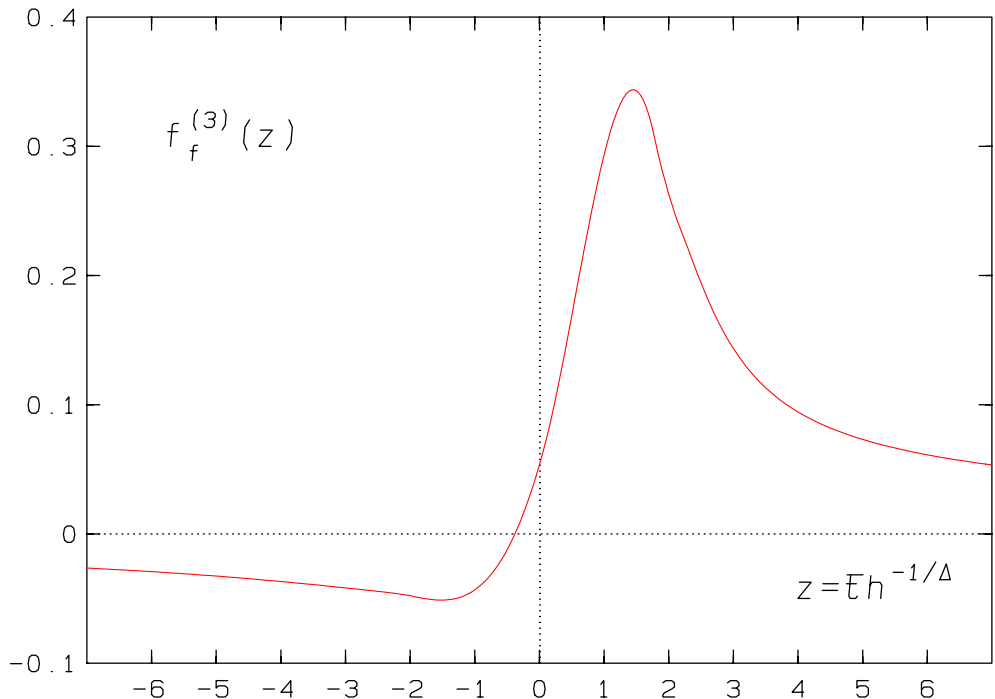, width=120mm}
\end{center}
\caption{The scaling function $f_f^{(3)}(z)$ as a function of 
$z=\bar t h^{-1/\Delta}$. The line shows our parametrization.}
\label{fig:ff3}
\end{figure}

Finally we show in Fig.\ \ref{fig:ff3} the third derivative of the scaling
function $f_f(z)$ with respect to $z$. It controls the singular behaviour
of the third derivative of the free energy density with respect to 
temperature,
\begin{eqnarray}
\frac{\partial^3 f}{\partial T^3} &\sim& \frac{\partial^3 f_{s}}
{\partial T^3}
= \frac{H_0}{T_0^3} h^{-(1+\alpha)/\Delta} f_f''' (z) \; ,
\label{f3}
\end{eqnarray}
and is the first thermal derivative of the free energy density that  
diverges at $T_c$ in the limit $H\rightarrow 0$. It thus allows a 
discussion of critical behaviour resulting from
\n the structure of the singular part of the free energy density without
the need of determining non-singular contributions to the free energy
density. This property of the three-dimensional $O(4)$ model has been
exploited in the discussion of critical behaviour in the vicinity of the
chiral phase transition of 2-flavor QCD \cite{Asakawa:2009aj,Friman:2011pf}.
\section{Summary and conclusions}
\label{section:conclusion}
In this paper we have investigated the scaling functions of the 
three-dimensional $O(4)$ model, which can be derived from the singular
part of the free energy density.  In contrast to other papers 
\cite{Engels:1999wf}-\cite{Braun:2007td} where the scaling functions 
are parametrized in the Widom-Griffiths form, we have chosen a form
which is preferred in tests of other models such as QCD. Here the 
scaling variable $z$ is independent of the magnetization. These scaling
functions can of course be derived from those of the other type, however
the explicit functional dependence on $z$ is of value. 
The major advantages of our parametrization are 
\begin{itemize}
\item[1)] derivatives with respect to the variable $z$ can be taken
   directly and not implicitly, the positions of extrema are 
   easily calculable.   
\item[2)] the parameters are determined from direct fits to the data, 
   one can immediately judge how well the data are represented. 
\end{itemize}
\n In order to carry out the parametrization we have used the best
presently available data set for finite magnetic fields
\cite{Engels:2009tv}. Furthermore, 
we have tackled the problem of the scaling of the energy density
and the specific heat in the $3d$ $O(4)$ model, to our knowledge for 
the first time at all, and we were able to clarify the r\^ole of
the non-singular part of the energy density for scaling. This is
very important for all corresponding checks of QCD with two light 
flavours (see for example Ref.\ \cite{D'Elia:2005bv}). Our approach
in some more detail was the following:

\n We have parametrized the scaling function $f_G(z)$ of the
magnetization with asymptotic expansions for $z \rightarrow \pm\infty\,$  
and Taylor expansions around $z=0$. The knowledge of the expansion
coefficients of $f_G(z)$ enabled us to derive the corresponding 
coefficients for the scaling function $f_f(z)$ of the singular part of
the free energy density. In particular, we could calculate the leading
asymptotic coeffcients $c_0^\pm$ of $f_f(z)\,$ and
thereby determine the universal amplitude ratio for the specific heat
to $A^+/A^-=1.842(43)\,$. In the following we have tested our data for
the energy density and the specific heat with the respective scaling
functions. To this end we have determined the non-singular parts of the 
two observables at $T_c\,$. Whereas this is always possible for the
energy density, we could do that for the specific heat only, because
in the three-dimensional $O(4)$ model the exponent $\alpha$ is negative.
With these results we found scaling for the energy density in the 
neighbourhood of $T_c$ or small $|z|$ but not outside. As it turned out, 
we can achieve scaling for all our $T$ or $J$-values, if we use the 
correct values of $\epsilon_{ns}(T)$. The latter can be calculated from
our function $f_f\p(z)$ if the data show scaling in $h$. We found 
indeed perfect scaling in $h$ and very accurate
$\epsilon_{ns}(T)$-values. They form a smooth function of $T$ in 
accord with the results from $T_c\,$. As expected, $T_c$ is not a
distinguished point of this function. For other models the regular 
parts of the energy density are of course different and may be not so
easy calculable, if for example correction-to-scaling terms spoil 
$h$-scaling. A test of the energy density and/or specific heat on 
scaling may then be problematic.

As a last result of our parametrization we quote the peak positions
of scaling functions which are relevant for the definition of 
pseudocritical lines. In all cases the peaks are very flat and symmetric
around the peak positions $z_p$. We found the value $z_p^{0,2}=1.374(30)$
for $f_{\chi}$, in $-f_G\p$ the value $z_p^{1,1}=0.74(4)$ and for
$-f_f''$ the value $z_p^{2,0}=-0.38(8)$. The corresponding observables
are $\chi_L$, $\chi_t$, and $C-2\epsilon/T$, the upper index of $z_p$
denotes the number of derivatives of the free energy density with respect
to $T$ and $H$. Evidently, the pseudocritical lines can be rather 
different. If, for example, the peak position of $\chi_t$ is used to 
define the pseudocritical temperature, it will be closer to $T_c$ as in
the case of $\chi_L$.  

\vskip 0.2truecm
\noindent{\Large{\bf Acknowledgments}}

\n This work has been supported in part by contracts DE-AC02-98CH10886
with the U.S. Department of Energy, the BMBF under grant 06BI401
and the Deutsche Forschungsgemeinschaft under grant GRK 881.

\clearpage

\end{document}